# Lattice and magnetic instabilities in $CaFe_2As_2$: A single crystal neutron diffraction study

A.I. Goldman[1,2], D.N. Argyriou[3], B. Ouladdiaf[4], T. Chatterji[5], A. Kreyssig[1,2], S. Nandi[1,2], N. Ni[1,2], S. L. Bud'ko[1,2], P.C. Canfield[1,2] and R. J. McQueeney[1,2]

[1] Department of Physics and Astronomy, Iowa State University, Ames IA 50011
[2] Ames Laboratory, US DOE, Iowa State University, Ames IA 50011
[3] Helmholtz-Zentrum Berlin für Materialien und Energie, 100 Glienicker str, D-14109 Berlin, Germany
[4] Institute Laue-Langevin, 38042 Grenoble Cedex, France
[5] Forschungszentrum Jülich Outstation at Institute Laue-Langevin, 38042 Grenoble Cedex, France

## Abstract

Neutron diffraction measurements of a high quality single crystal of $CaFe_2As_2$ are reported. A sharp transition was observed between the high temperature tetragonal and low temperature orthorhombic structures at $T_S$ = 172.5K (on cooling) and 173.5K (on warming). Coincident with the structural transition we observe a rapid, but apparently continuous, ordering of the Fe moments, in a commensurate antiferromagnetic structure is observed, with a saturated moment of $0.80(5)\mu_B$/Fe directed along the orthorhombic a-axis. The hysteresis of the structural transition is 1K between cooling and warming and is consistent with previous thermodynamic, transport and single crystal x-ray studies. The temperature onset of magnetic ordering shifts rigidly with the structural transition providing the clearest evidence to date of the coupling between the structural and magnetic transitions in this material and the broader class of iron arsenides.

The excitement generated by the discovery of superconductivity at temperatures above 30K in the doped iron arsenide compounds[1,2,3,4] continues to spur further research around the world. Much of this work is directed towards elucidating the underlying pairing mechanism and its relationship to anomalies in thermodynamic and transport data associated with structural and/or magnetic transitions between 100K and 200K found in the "parent compounds," RFeAsO (R= Rare earth) and $AFe_2As_2$ (A=Ba, Sr, Ca).[4,5,6,7,8,9] An apparent prerequisite for superconductivity in both classes of iron arsenides is the elimination of these higher temperature transitions through doping with fluorine for oxygen in the former compound, and potassium or sodium for the A ion in the latter. The introduction of new chemical species, however, adds further complexity since a range of effects (e.g. chemical disorder, doping inhomogeneities) must be considered. The recent discoveries[10] that, in $CaFe_2As_2$: (i) under modest hydrostatic pressure, the first-order structural phase transition can be suppressed; (ii) over a limited pressure range, superconductivity occurs and; (iii) at higher pressures superconductivity can again be suppressed with the stabilization of a potentially different high temperature phase, provides a valuable new parameter for tuning the behavior of these fascinating compounds. Further, these observations support the notion of a strong interaction between lattice and magnetic degrees of freedom and superconductivity in the $AFe_2As_2$ family. For example, strong spin-fluctuations due to magnetic frustration in the doped compounds may play an important role in the superconductivity. The importance of frustration as a mechanism for driving structural transformations in the parent phases and supporting spin fluctuations in the superconducting phases, has been discussed in several papers.[11,12] In addition, Fermi surface nesting features apparent from band structure calculations may give rise magneto-elastic coupling leading to charge/spin-density wave transitions.[13]

Several groups have used powder and/or single crystal diffraction to study the structural transition in $BaFe_2As_2$,[8,14,15] $SrFe_2As_2$,[16,17,18] $EuFe_2As_2$,[18] and $CaFe_2As_2$,[19] which is quite similar in nature to the corresponding structural transformation observed in the un-doped RFeAsO compounds. For the $AFe_2As_2$ series, the high temperature tetragonal (I4/*mmm*) structure transforms to an orthorhombic (F*mmm*) unit cell rotated 45° with respect to the tetragonal basal plane axes. The structural transition appears discontinuous and often hysteretic, and coexistence is generally observed between the high temperature tetragonal phase and the lower symmetry orthorhombic phase close to the structural transformation; all hallmarks of a first-order transition. In all of the doped, superconducting $AFe_2As_2$ phases to date, this structural transformation is suppressed or absent. Studies of the magnetic ordering that arises in the vicinity of the structural transformation in the un-doped iron arsenides have been accomplished using neutron powder diffraction and Mössbauer measurements. In contrast to the behavior observed for the un-doped iron oxypnictides[20] powder neutron diffraction from $BaFe_2As_2$[14] and $SrFe_2As_2$[17] (together with Mössbauer measurements) show that the magnetic ordering appears to be concomitant with the structural distortion. The magnetic ordering is signaled by a sharp, discontinuous transition for $SrFe_2As_2$, as measured by Mössbauer spectroscopy. For $BaFe_2As_2$, however, the magnetic order parameter measured by powder neutron diffraction appears nearly continuous. Although the coincidence in temperature of the structural and magnetic transitions argues strongly for coupling between the lattice and spin degrees of freedom in these phases, further work intent on clarifying the nature of these transitions and the relationship between them is required to develop a full picture.

In particular, previous neutron scattering measurements of the structural and magnetic anomalies in these compounds have been based on powder diffraction, rather than single crystal studies, leading to some ambiguities in the details of the magnetic structure and moment direction.[14] For the lower symmetry orthorhombic structure single crystal studies allow an unambiguous determination of the magnetic structure, including the direction of the magnetic moment and the magnetic propagation vector with respect to the orthorhombic unit cell. Investigations of the structural and magnetic transitions in $CaFe_2As_2$, particularly in light of its novel behavior under pressure, can provide important new insight into the nature of the structural and magnetic transitions.

In this Communication we describe neutron diffraction measurements of a high quality single crystal of $CaFe_2As_2$. A sharp transition was observed between the tetragonal and orthorhombic structures at $T_S$ 172.5K (on cooling) and 173.5K (on warming). The hysteresis of ≈ 1K between cooling and warming for the structural transition is consistent with previous thermodynamic and single crystal x-ray studies.[19] Coincident with the transition from the high temperature tetragonal to the low temperature orthorhombic phase we observe a rapid, but continuous ordering of the Fe moments, in a commensurate antiferromagnetic structure, that saturates at a value $0.80(5)\mu_B$/Fe directed along the longer orthorhombic $a$-axis. The onset of magnetic ordering shifts rigidly with the structural transition providing the clearest evidence to date of the coupling between the structural and magnetic transitions in this family of iron arsenides.

For the neutron diffraction measurements, single crystals of $CaFe_2As_2$ were grown out of a Sn flux using conventional high temperature solution growth techniques described previously.[19] A 7 mg single crystal with dimensions of approximately 3mm X 2mm X 0.2mm was selected for the neutron diffraction measurements. X-ray and neutron Laue measurements confirmed the quality of the sample (mosaic ≈ 0.5 degrees FWHM) and that the tetragonal $c$-axis is perpendicular to the flat plate-shaped crystal surface. The neutron diffraction experiment was performed on the four-circle station, D10, at the Institute Laue-Langevin (ILL) using a wavelength of 2.36Å and a two-dimensional area detector mounted on the scattering arm for data collection. A pyrolitic graphite filter was employed to reduce the higher harmonic content to less that $10^{-4}$ of the primary beam energy. The sample was mounted on the cold-finger of a He flow cryostat which provided a temperature stability of 0.1K.

We first describe the chemical and magnetic structures above and below the transition temperature, $T_S$. For temperatures above $T_S$ ≈ 173K, the crystal structure of $CaFe_2As_2$ is well described by the same tetragonal structure as its sister compounds, $SrFe_2As_2$ and $BaFe_2As_2$, but with lattice constants $a_{Tet}$ = 3.879(3) Å and $c_{Tet}$ = 11.740(3) Å at T=300 K. Figure 1(a) shows longitudinal scans taken through the allowed nuclear (1 -1 -2) reflection for the tetragonal structure. For reference, Fig. 1(a) also shows a scan through the position where the magnetic peak is found in the low-temperature orthorhombic phase. As the sample was cooled below $T_S$, Figure 1(b) shows that the nuclear peak appears to split, signaling the transition to the orthorhombic phase with $a_{Orth}$ = 5.506(2) Å , $b_{Orth}$ = 5.450(2) Å and $c_{Orth}$ = 11.664(6)Å at T = 10K. For the orthorhombic structure we employ indices (H K L) for the reflections based on the relations: H = h – k, K = h + k and L = l, where (h k l) are the corresponding Miller indices for the tetragonal phase. For example, the (1 -1 -2) tetragonal peak is properly labeled as the orthorhombic (2 0 -2) peak below the structural transition. The other peak, (0 2 -2)′, evident in Figure 1(b) arises from the same twinning processes previously described in detail for $YBa_2Cu_3O_{7-\delta}$[21]. Indeed, two, three or four reflections associated with twinning are observed at the nuclear positions depending on the setting of the scattering plane relative to the resolution function of the instrument.[22] However, for the lower symmetry of the orthorhombic cell, twinning produces domains along a given direction that are not equivalent and, typically, the magnetic propagation vector is aligned along a unique direction in each of the twin domains. The bottom panel of Figure 1(b) plots all of these peaks as a function of both Q and ω (sample rotation angle in the scattering plane), to further clarify the relationship between the twinned domains and the magnetic peak (the Q-scans displayed above these plots represent a projection of the Q-ω plane on the Q-axis). By comparing the position of the magnetic peak on the right side of Fig 1(b), with the positions of the structural peaks on the left side of Fig. 1(b) we see that the magnetic peak is properly labeled as the (1 0 -1) magnetic reflection. More specifically, the magnetic peak appears at a position that is at half the wavevector, and the same ω, as the (2 0 -2) nuclear peak.

At low temperature (T=10K) a set of nuclear reflections and magnetic peaks were collected for refinement of the chemical and magnetic structure using FULPROF. The chemical unit cell below $T_S$ is well described by the orthorhombic structure F*mmm* with $z_{As}$ = 0.36642(5) for As in the 4e sites. Refinement of the magnetic structure with a magnetic propagation vector of (0 1 0)[23] yielded an ordered Fe moment of 0.80(5)$\mu_B$/Fe directed along the longer orthorhombic *a*-axis. The size and direction of the Fe moment was determined from a least squares fit of 33 magnetic reflections giving an acceptable fit (R=8.8%) in light of the size of the crystal and magnetic moment. A refinement of the magnetic structure using a model with the Fe moments directed along the shorter *b*-axis, provided poor fits (R=84%) to our single crystal data. The results of the magnetic structural refinement for the magnetic structure and moment direction of $CaFe_2As_2$ are illustrated in Figure 2. The magnetic cell is the same as the chemical unit cell and the moments are ferromagnetically coupled along the *b*-direction and antiferromagnetically coupled along the *a*- and *c*-directions.

The most interesting aspect of these measurements deals with the behavior of the magnetic and structural order close to the transition temperature, $T_S$. The temperature profiles of both the tetragonal to orthorhombic transition and the magnetic ordering are illustrated in Figures 3(a). These particular data were obtained on warming the sample from base temperature through $T_S$ and indicate the temperature evolution of longitudinal Q-scans through the (0 4 0)/(4 0 0) twins (in orthorhombic notation) and the (-1 0 1) magnetic peak. Above $T_S$, only the (2 2 0) tetragonal nuclear peak is observed. From these panels, a rather abrupt transformation in both the crystal structure and magnetic ordering at $T_S$ might be inferred.  Below approximately 170K, both the orthorhombic distortion and the magnetic peak intensity are essentially saturated.  More detail of the behavior in the transition region itself is provided in Fig. 3(b) which plots the temperature dependence of both the orthorhombic splitting of the nuclear peak and the intensity of the magnetic peak upon warming and cooling through the transition.  Here we see that while the structural transition, as measured by the orthorhombic splitting, is discontinuous over a temperature range of less than 0.5K, the magnetic ordering appears to evolve continuously over a short, but finite temperature range.  In this transition region, where both the tetragonal and orthorhombic phases coexist, the volume fraction of the orthorhombic phase also changes making it difficult to clearly identify the order of the magnetic transition in the present measurement. Nevertheless, it is clear that the onset of magnetic ordering occurs just at $T_S$.  Figure 3(b) also displays a clear signature of hysteresis in the tetragonal-to-orthorhombic transformation over a range of approximately 1K. Perhaps most interesting, however, is that the magnetic ordering shifts rigidly with the structural transition. This provides the clearest demonstration, to date, that the structural transition and magnetic ordering are intimately connected.

It is useful at this point to compare these results to those for the magnetic structures of $SrFe_2As_2$ and $BaFe_2As_2$.  Whereas the magnetic ordering in $CaFe_2As_2$ is consistent with that proposed for $BaFe_2As_2$, the powder measurements on the latter system could provide only a representational solution of the magnetic structure rather than a unique ordering scheme (propagation vector and moment direction) with respect to the orthorhombic unit cell.[14]   Therefore, it remains possible that the details of the magnetic structure for $BaFe_2As_2$ are different from what we have found for $CaFe_2As_2$. Interestingly, the structural transition for $BaFe_2As_2$ appears to be first order and the magnetic ordering appears to be continuous, in agreement with the trends observed in our measurements although the error bars on the powder data preclude a definitive conclusion for $BaFe_2As_2$.  The magnetic structure determined from high resolution magnetic powder data for $SrFe_2As_2$ by Jesche et al,[17] is consistent with the present single crystal measurements.  Indeed, the ordered moments determined for all three members of the $AFe_2As_2$ (A = Ba, Sr, Ca) are closely similar and larger than those determined for Fe in the RFeAsO (R= Rare Earth) compounds.   However, the data for magnetic ordering in $SrFe_2As_2$ also clearly indicate that both the structural and magnetic ordering are discontinuous at the transition; both appear to be first order in nature.

The results presented here clarify two important issues for $CaFe_2As_2$ in particular and the $AFe_2As_2$ family of compounds in general:  (i) they elucidate the complex nature of the ≈173K phase transition seen at ambient pressure in $CaFe_2As_2$ and, (ii) they provide a fixed point for speculation about the effects of pressure on this intriguing compound.  Whereas it is true that the proximity of the magnetic and structural transitions supports the notion of coupling between the transitions, the fact that we observe (Fig. 3b) the magnetic ordering rising just below the structural phase transition, even when the first order structural phase transition is shifted by thermal hysteresis, unambiguously links these two phase transitions at ambient pressure.  The magnetic and structural phase transitions are, indeed, strongly coupled in $CaFe_2As_2$.  This observation of such strong coupling, though, does bring up the bedeviling question of which transition, magnetic or structural, is driving the other. As shown in ref. 10, when hydrostatic pressure is applied, the signature of the high temperature phase transition becomes more second order like and finally disappears near 5 kbar, close to the pressure about which a superconducting dome is centered for T < 12 K. For even higher pressures a different resistive anomaly, one more commonly associated with a loss of spin disorder scattering and the onset of antiferromagnetism without any gapping of the Fermi surface, is observed. We anticipate that neutron scattering on $CaFe_2As_2$ under hydrostatic pressure will further clarify the relationship between these transitions as it appears likely that for pressures larger than 5 kbar at least one of them (most likely the structural one) will be fully suppressed.

**Note added:** Upon completion of this work we became aware of a single crystal neutron diffraction measurement on $SrFe_2As_2$ which finds the same magnetic structure and ordered moment as we described here.[24]

**Acknowledgment**

Work at the Ames Laboratory was supported by the US Department of Energy – Basic Energy Sciences under Contract No. DE-AC02-07CH11358. We gratefully acknowledge the ILL for the rapid allocation of time and their support for this work. The authors would like to thank G. McIntyre for his assistance in identifying the twinning scheme.

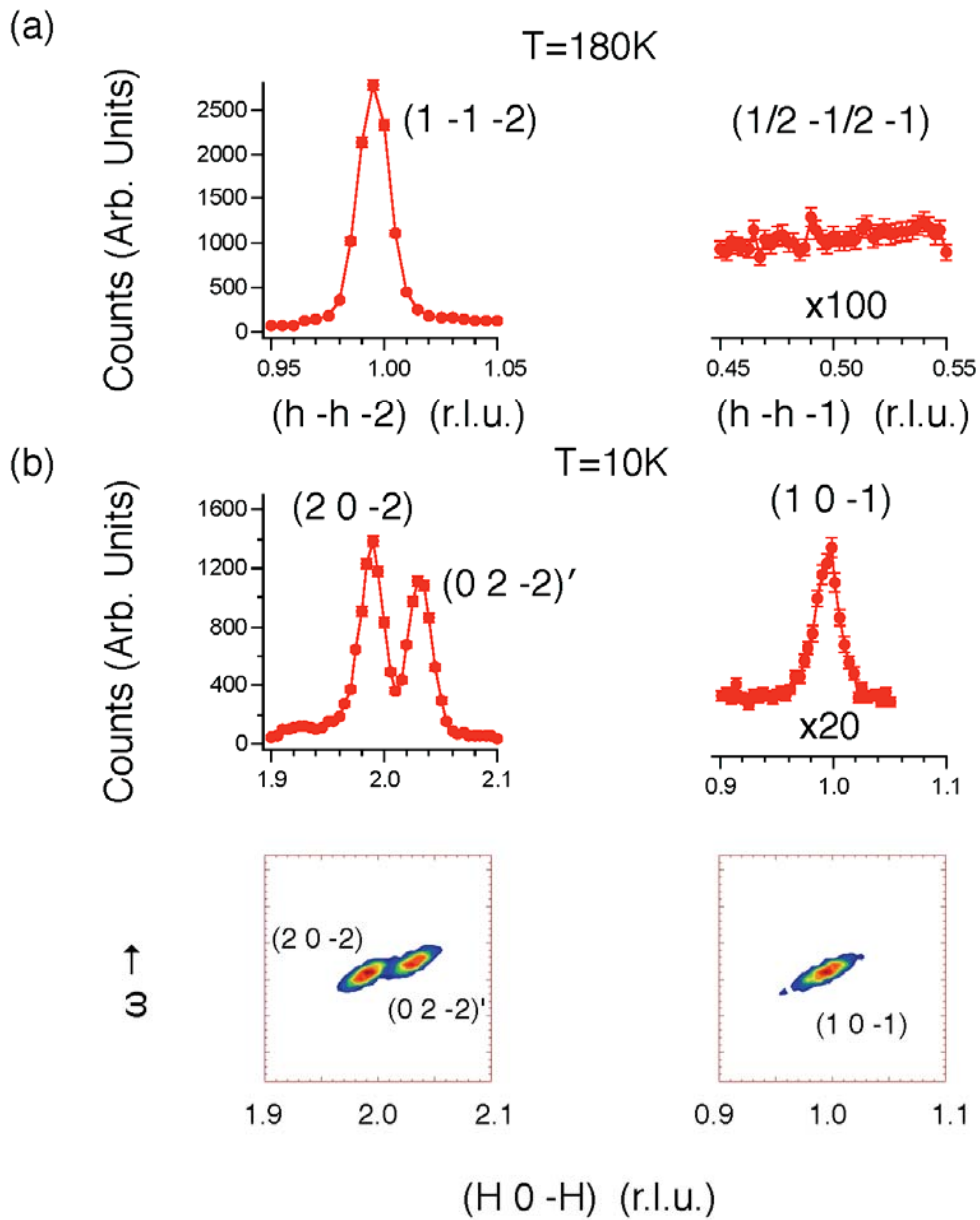

**Figure 1. (a)** Q scans through the positions of the tetragonal (1 -1 -2) nuclear peak (referenced to the tetragonal unit cell) and the magnetic peak positions at 180K. No intensity is observed at the positions of the magnetic reflections. **(b) (top)** Below the tetragonal-to-orthorhombic transition, two twin domains are observed in longitudinal scans through the (2 0 -2) (indexed to the orthorhombic cell) nuclear peak position. The magnetic peak, (1 0 -1), is associated with only one of these domains. **(bottom)** Two-dimensional plots showing the Q and ω (sample rotation in the scattering plane) dependence of peaks in the top panel.

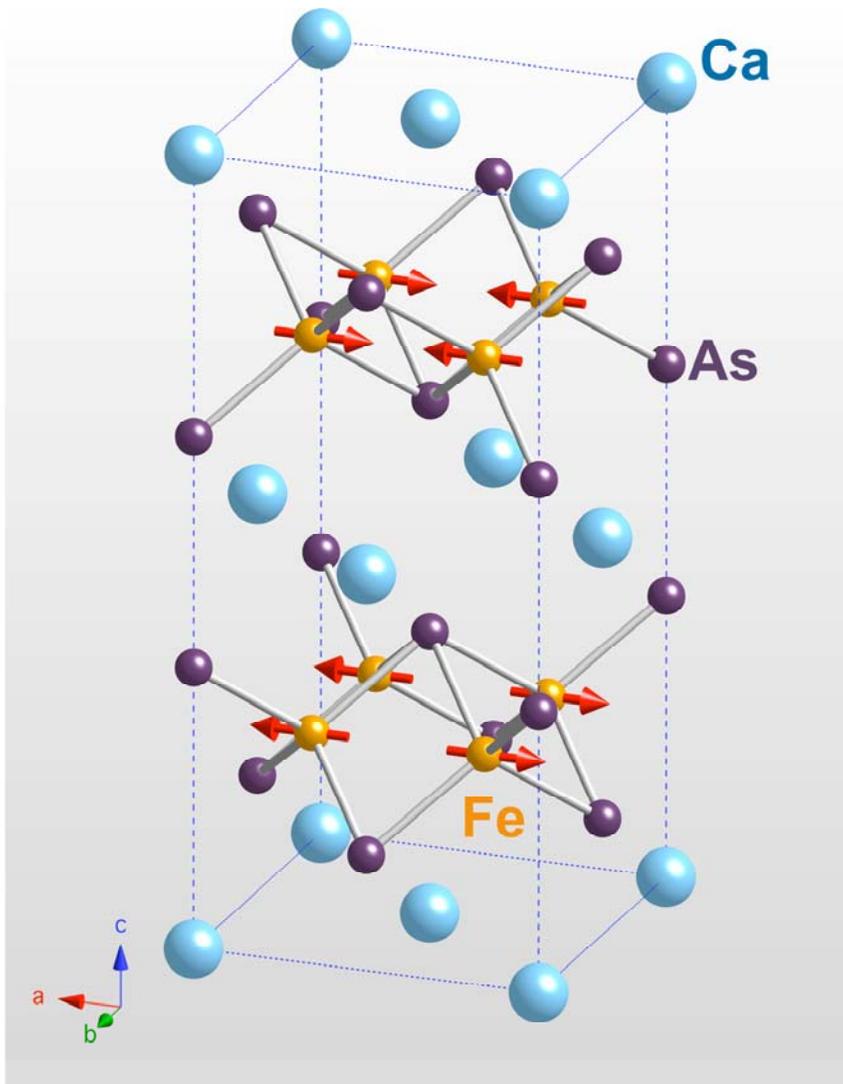

**Figure 2.** Illustration of the antiferromagnetic structure of $CaFe_2As_2$ below $T_S$. The magnetic unit cell is the same as the orthorhombic chemical unit cell. Fe moments are oriented along the orthorhombic *a*-axis.

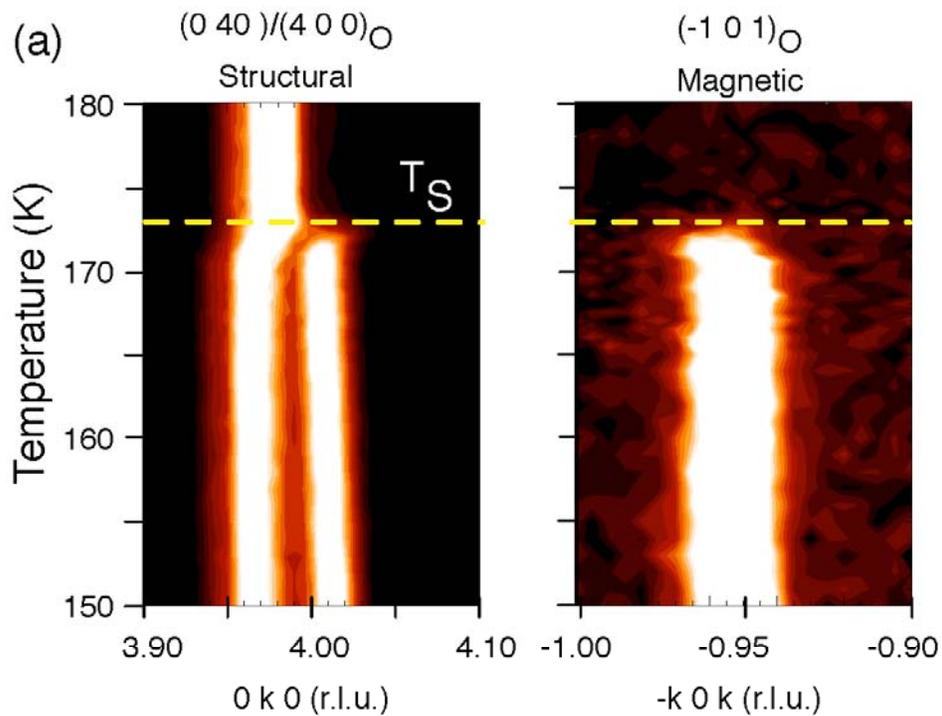

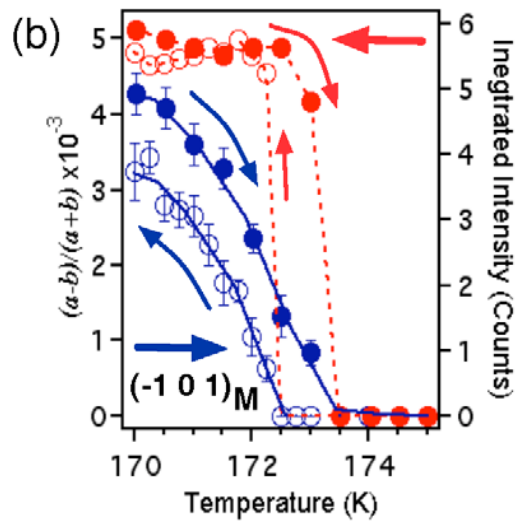

**Figure 3.** (a) Plot of the temperature dependence of Q-scans through the (0 4 0)/(4 0 0) twins [referenced to the orthorhombic cell] in the vicinity of the tetragonal-to-orthorhombic structural transition at $T_S$ = 173.5K (on warming). Above the transition, the single peak is the tetragonal (2 2 0) (b) Temperature dependence, close to $T_S$, of the orthorhombic splitting (red curves) and magnetic integrated intensity of the (-1 0 1) reflection (blue curves) upon warming (open circles) and cooling (closed circles) through the transition. Below 170K, both the orthorhombic splitting and the magnetic peak intensity are saturated.